\documentclass[a4paper,11pt]{article}
\usepackage{amsmath}
\usepackage{amsfonts}
\usepackage{amssymb}
\usepackage{bm}
\usepackage[utf8x]{inputenc}
\usepackage{ae}
\usepackage[T1]{fontenc}
\usepackage[english]{babel}
\usepackage{graphicx}
\usepackage{makecell}
\usepackage{placeins}
\usepackage{tabularx}
\usepackage{tabulary}
\usepackage{setspace}
\usepackage{tablefootnote}
\usepackage[flushmargin]{footmisc}
\usepackage[comma]{natbib}
\usepackage{array}
\usepackage{float}
\usepackage{caption}
\usepackage{tikz}
\usetikzlibrary{shapes.geometric}
\usepackage[a4paper,left=2.5cm,right=2.5cm,top=3cm,bottom=3cm]{geometry}
\usepackage[colorinlistoftodos]{todonotes}
\usepackage{pdfpages}
\usepackage{longtable}
\usepackage{adjustbox}
\usepackage{booktabs}
\usepackage{multirow}
\usepackage{lscape}
\usepackage[colorlinks=true, allcolors=blue]{hyperref}

\usepackage[flushleft]{threeparttable}
\usepackage{inputenc}



\begin{document} \doublespacing \pagestyle{plain}
	
	\def\ci{\perp\!\!\!\perp}
	\begin{center}
		
		{\LARGE On suspicious tracks: machine-learning based approaches to detect cartels in railway-infrastructure procurement}

		{\large \vspace{0.8cm}}
		
		{\large Hannes Wallimann and Silvio Sticher }\medskip

		{\small {University of Applied Sciences and Arts Lucerne, Institute of Tourism and Mobility} \bigskip }
		
		{\large \vspace{0.8cm}}
		
		{\large Version March 2023}\medskip

	\end{center}
	
	\smallskip

	\noindent \textbf{Abstract:} {
		In railway infrastructure, construction and maintenance is typically procured using competitive procedures such as auctions.
		However, these procedures only fulfill their purpose---using (taxpayers’) money efficiently---if bidders do not collude. Employing a unique dataset of the Swiss Federal Railways, we present two methods in order to detect potential collusion: First, we apply machine learning to screen tender databases for suspicious patterns. Second, we establish a novel category-managers' tool, which allows for sequential and decentralized screening. To the best of our knowledge, we pioneer illustrating the adaption and application of machine-learning based price screens to a railway-infrastructure market. 	
	}
	
	{\small \smallskip }
	{\small \smallskip }
	{\small \smallskip }
	
	{\small \noindent \textbf{Keywords:} railway infrastructure, cartel detection, machine learning, procurement auctions}
	
	{\small \smallskip }
	{\small \smallskip }
	{\small \smallskip }
	
	{\small \noindent \textbf{Acknowledgments:}  Ramon Hauser, Lea Oberholzer, and Alisha Bühler provided excellent research assistance. Moreover, the authors would like to thank Philipp Wegelin, Karin Amberg, Marco Fetz, and Zeno van Uden for helpful comments. Finally, we are grateful to the SBB Research Fund for financial support. All the views expressed in this paper are solely those of the authors. }
	
	\bigskip
	\bigskip
	\bigskip
	\bigskip
	
	{\small {\scriptsize 
\begin{spacing}{1.5}\noindent  
\textbf{Addresses for correspondence:} Hannes Wallimann, University of Applied Sciences and Arts Lucerne, Rösslimatte 48, 6002 Lucerne, \href{mailto:hannes.wallimann@hslu.ch}{hannes.wallimann@hslu.ch}; Silvio Sticher, \href{mailto:silvio.sticher@hslu.ch}{silvio.sticher@hslu.ch}.
\end{spacing}
			
		}\thispagestyle{empty}\pagebreak  }

	{\small \renewcommand{\thefootnote}{\arabic{footnote}} %
		\setcounter{footnote}{0}  \pagebreak \setcounter{footnote}{0} \pagebreak %
		\setcounter{page}{1} }
	
\section{Introduction}\label{introduction}

According to recent OECD data, investment in railway infrastructure in 2019 by 23 of the 27 countries of the European Union amounted to EUR 38,3 bn.\footnote{See \hyperlink{https://data.oecd.org/transport/infrastructure-investment.htm}{https://data.oecd.org/transport/infrastructure-investment.htm} (accessed on September 19, 2022)} France (EUR 10,8 bn) and Germany (EUR 9.6 bn) alone are responsible for around half of this amount \citep{ITF}. There are several reasons why investment in railway infrastructure (mainly construction but also maintenance) could increase in the near future. For example, the European Union has established a new action plan to boost long-distance and cross-border passenger rail to foster sustainable mobility \citep{EuropeanComission2021}. Furthermore, the trend towards intermodal and multi-modal mobility requires corresponding facilities such as transport hubs and IT systems such as digital platforms \citep{montero2020railway}.
	
 Public transportation in general and railways in particular exhibit a great extent of scale economies. This renders railway infrastructure a natural monopoly \citep{crozet2012beyond}. Usually, infrastructure is owned and managed by a state-owned infrastructure company or a separate infrastructure division of a state-owned transport company \citep{finger2014governance}. The accordingly substantial public subsidies involved \citep[see, e.g., ][]{finger2014governance,montero2020railway} raise the need for means to ensure that the taxpayers' money is used efficiently. In the wake of liberalization of the public sector in general and of the railway sector in particular, many infrastructure companies were compelled to start procuring construction and maintenance using competitive procurement procedures such as auctions \citep{laffont1993theory}. 

This is supposed to ensure efficient use of (partially) public money. However, as common to procurement auctions, concerns regarding bid-rigging appear \citep{lengwiler2006corruption}. Therefore, and because of the large amount of public money involved, infrastructure companies and their regulators have a vital interest in preventing unlawful cooperation of firms applying for projects. Detecting collusive arrangements is an indispensable first step in fighting cartels. In order not to rely heavily on external sources---such as principal witnesses in leniency programs---researchers have proposed statistical methods to screen markets to detect possible patterns of a cartel \citep[see, e.g., ][]{Porter1993,Harrington2008}. 

To the best of our knowledge, none of the existing studies focus on approaches designed for detecting potential cartel behavior in the railway-infrastructure industry. We aim to fill this gap by presenting two complementary approaches to detect potential illegal supplier cartels. Both of these approaches are related to cartel screening using machine learning \citep[for a review see, e.g., ][]{harrington2022cartel}, where algorithms "learn" to predict the presence or absence of a cartel. First, we directly apply machine-learning based algorithms to a pooled database with many projects previously put to tender (centralized approach). Second, we showcase a tool created specifically for category managers of railway-infrastructure companies (decentralized approach). With the tool, a category manager can directly (live) monitor potential bid-rigging in his or her individual tenders by benchmarking descriptive statistics with critical values. Thereby, we use machine learning to select the relevant statistics and their critical values.

As an illustration for ex-ante cartel screening in the railway-infrastructure industry, we apply both approaches to the Swiss context by using unique procurement data provided by the Swiss Federal Railways, for which a cartel's incidence is unknown. The data set consists of 1,818 tenders, mainly from 2016 to 2021. We classify a (rather low) share of between 2.6\% and 10.9\% of all tenders as suspicious, depending on which approach and, more specifically, which decision threshold (i.e., probability threshold) of machine-learning algorithms we apply (0.5 and 0.7, respectively). We then present several further investigations of the "suspicious" tenders, an exercise of potentially direct practical use, as railway-infrastructure companies can use our investigation steps to explore their own markets. Note that, before applying either approach, we develop models using past data from investigations by the Swiss Competition Commission \citep[see also ][]{wallimann2022machine,Imhof2019}. We find that we correctly classify a tender in 80\% of the cases. Furthermore, the 95\% prediction interval for correct classification (potential collusion or no collusion, respectively) reaches from 70\% to 90\%. 

We proceed as follows. In Section \ref{Litrev}, we present the relevant literature on screening methods. In Section \ref{Toolbox}, we discuss our cartel-detection approaches and their application to railway infrastructure in general. In Section \ref{Fitmodels}, we adapt them to the Swiss railway-infrastructure market. In Section \ref{Application}, we screen this market for cartel-like bidding patterns using unique procurement data provided by the Swiss Federal Railways. In Section \ref{Discussion}, we discuss our results and argue that the external validity is likely to be given---i.e., the market structures in the cases from the Swiss Competition Commission and our application are comparable. In Section \ref{Conclusion}, we conclude.

\section{Literature review on screening methods}\label{Litrev}

Our work is related to studies discussing pro-active statistical methods to detect potential cartels, initially proposed by, for instance, \citet{Harrington2008} and \citet{Porter1993}. These so-called screening methods constitute the first step of a multi-phase process that finally may condemn illegally cooperating firms. From an empirical perspective, screening methods are applied in the presence of "prediction policy problems" \citep{kleinberg2015prediction}. That is, researchers aim to predict the probability of a cartel. As we use bidding patterns in procurement auctions to identify potential cartels, we focus on price screens in this review section. Moreover, our main attention is on behavioral screens, observing firms' behavior in markets. This is opposed to structural screens, which identify markets with typical cartel-conducive traits such as a small number of bidders \citep{Harrington2008}. Note that the German Railways use structural screens to identify markets prone to cartels \citep[see, e.g., ][]{beth2022cartel}.

As we show in Section \ref{Fitmodels}, we find that the coefficient of variation, defined as the standard deviation divided by the arithmetic mean of all bids submitted in an auction, is highly predictive in terms of determining the presence of a potential cartel. This fits the body of literature that uses screens to analyze the variance of bids: Several studies \citep[e.g., ][]{Abrantes-Metz2006,Jimenez2012,Imhof2019} point out that the coefficient of variation decreases during periods in which a cartel is active. Other studies find that cartel-involved behavior is often accompanied by asymmetrical bidding distributions. Recent findings from Japan \citep{chassang2022robust}, Canada \citep{Clark2018}, and Switzerland \citep{Imhof2018} show that winning bids tend to be isolated; that is, the difference between the first- and second-lowest bids of a cartel is relatively high compared to differences among losing bids. 

In the past five years, the literature in the context of cartel screens has largely focused on the use of machine learning, a subfield of artificial intelligence \citep[][]{OECD2022}. It seems promising in light of the wide-scale use of digitization in procurement processes.  The corresponding literature includes papers by \citet{huber2019machine}, and \citet{silveira2022won} who apply machine learning algorithms (together with price screens) to find possible patterns of cartels in Switzerland, Croatia, and Brazil, respectively. Moreover, \citet{rodriguez2022collusion}, and \citet{imhof2021detecting} use machine learning algorithms in multiple countries. The latter paper proposes a method to identify conspicuous groups of firms directly instead of tenders affected by conspiracies. \citet{huber2022transnational} investigate the transnational transferability of machine-learning algorithms. They show that a country's institutional context matters for a cartel's influence on the distribution of bids. Therefore, applicants should use algorithms trained with data from other countries only with great reluctance. \citet{wallimann2022machine} detect incomplete cartels, where only some but not all bidders are members of the colluding agreement. Finally worth mentioning is the study of \citet{silveira2023you} using unsupervised machine learning for cartel detection. 
	
\section{Methods}\label{Toolbox}

We recommend two complementary approaches for railway-infrastructure companies to detect potential cartels and, thus, avoid paying excessively high prices. 
First, we show how machine-learning techniques can be implemented into the wide-scale use of digitization in the railways' infrastructure-procurement process (centralized approach). Thereby, screens are used as predictor variables to assess whether a cartel is likely to be present in a tender.
Second, we propose a tool specifically designed for category managers to classify tenders according to statistical benchmarks (decentralized approach). Here, we define the relevant benchmarks with an intuitive machine-learning algorithm, the classification tree.
Both of our approaches require minimal data: We only need the bids submitted by the firms in a procurement auction. This is valuable insofar as obtaining truthful information variables other than price (e.g., business-specific variables such as capacity utilization) is challenging without attracting the attention of a cartel.

Upon establishing the prediction framework we consistently use, we introduce both respective approaches in greater detail in the following. 

\subsection{Prediction framework}\label{MLandPred}

We use supervised machine learning with a set of predictor variables ($X$) to predict our outcome of interest ($Y$), the presence of a potential cartel. The outcome variable (also: target variable) is binary---taking on the value 0 if firms compete and the value 1 if they collude. To investigate the performances of machine-learning algorithms, we randomly divide the data set into two parts. The training set consists of 75\% of the observations, and we use it to develop predictive models observing both predictor variables and outcomes. The test sample consists of the remaining 25\% of observations and is used to evaluate model performance by predicting whether a potential cartel is active or not. Thereby, we can assess the goodness of fit of the models by comparing the predictions with the actual outcomes. To do so, we apply two measures: The correct-classification rate and the F1 score. The correct-classification rate denotes the proportion of correctly classified tenders in the test set. The F1 score is a harmonic mean of the precision and recall (also: sensitivity) in the test set. Precision calculates the proportion of how many cartel classifications are correct (true positives divided by all cases classified as "suspicious"), whereas recall measures the proportion of actual cartels that the algorithms correctly classify (true positives divided by all true cartel cases).

We exclusively use descriptive statistics of the bids in a tender as predictor variables. These screens describe bidders' behavior in a tender and capture distributional changes due to a cartel. These exist because cartel members (explicitly or tacitly) coordinate bids in a tender and are aware of their 's bids. For our application, we use nine frequently discussed screens \citep[see, e.g., ][]{rodriguez2022collusion,huber2022transnational,wallimann2022machine}: the coefficient of variation (CV), the spread (SPD), the difference between the two lowest bids (DIFFP), the relative distance (RD), the alternative relative distance (ALTRD), the normalized distance (NORMRD), the skewness (SKEW), and the Kolmogorov-Smirnov test (KSTEST). These screens are formally described in Appendix \ref{Appendix_A}.

\subsection{Centralized approach: An add-on to the digitization of the railways' infrastructure procurement}\label{Approach1}

\begin{figure}[H] 
	\includegraphics[scale=.4]{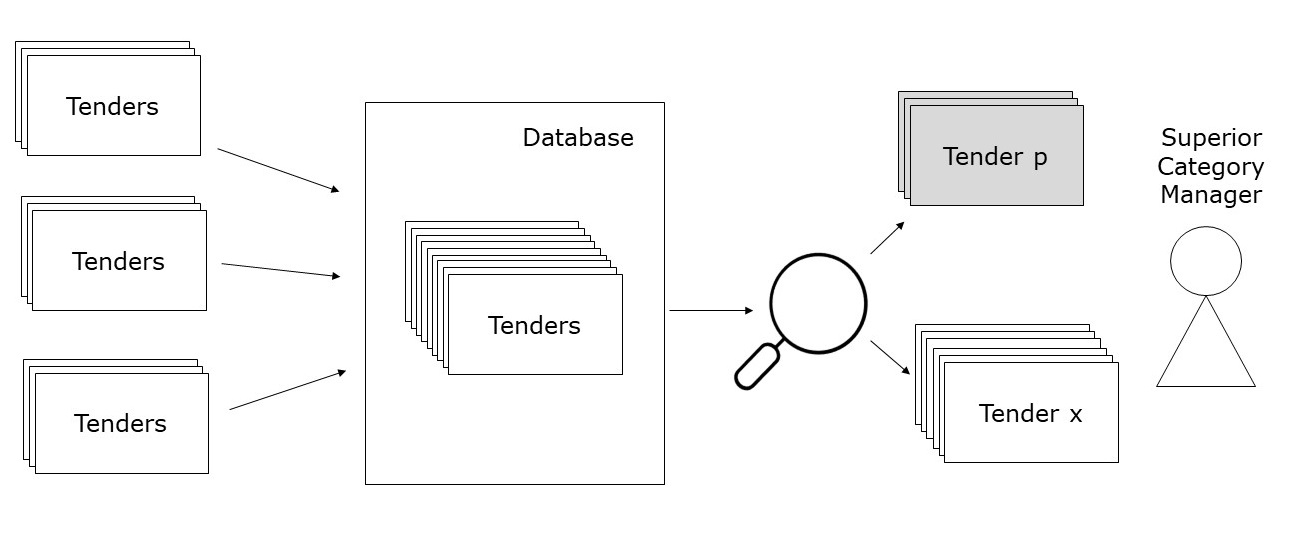}
	\centering \caption{Centralized approach} \label{Tool2_Fig}
\end{figure}

Our first approach (depicted in Figure \ref{Tool2_Fig}) relies on the instance of an increasingly digitized procurement, which allows to collect bids of firms in a single structured database (optimally in an automated way). Machine-learning algorithms then investigate the database with regard to potential cartels. They classify tenders as suspicious (the stack marked grey, including Tender $p$) or non-suspicious (the stack kept white, including Tender $x$). These evaluations are directly made available to the superior category manager of a railway-infrastructure company. 

As all tenders are pooled, the centralized (or cockpit) approach allows for both complete and disaggregated overviews of potentially suspicious clusters, i.e., broken down by regions, time periods, auction formats, and firms involved. 

Note that until now, we simplified the classification task by predicting a binary target taking the values 0 or 1, i.e., whether firms collude or not. However, what we really do is carrying out so-called probability predictions. The intuitive default value for machine-learning algorithms is 0.5: If a tender is assigned a probability of 0.5 or higher, the algorithm classifies the outcome value to 1, i.e., "suspicious" in terms of potential cartel participation. Stated differently, this rule puts each tender to the class that is more likely. Increasing the threshold, e.g., to 0.7, allows applicants (i.e., railway-infrastructure companies) to avoid false-positive results, i.e., wrongly classifying bidders as a cartel. On the other hand, however, it increases the occurrence of false-negative results, i.e., a cartel is wrongly discarded. In the literature, different thresholds are proposed. \citet{huber2019machine}, for instance, suggest a decision threshold between 0.5 and 0.7, while \citet{silveira2022won} discuss a threshold between 0.6 and 0.75. We combine these suggestions by labeling tenders with a probability prediction between 0.5 and 0.7 as "suspicious" and those with an even higher value as "very suspicious". 

Rail infrastructure companies could use a variety of algorithms to detect potential cartels using this approach, three of which we present in the following. The first is a logistic-regression model \citep{berkson1944application}. The second algorithm is the random forest \citep{Breiman2001} which builds on classification trees. The third algorithm is the super learner \citep{vanderLaanetal2007}, which is an ensembled method averaging several other machine learning algorithms, i.e., random forest, lasso logit regression, gradient boosting, and neural networks. 

First, we implement the (binary) logistic regression as baseline model in order to highlight the contribution to the literature by (newer) machine-learning algorithms. The logistic regression has a binary outcome variable. In our case, it takes on the values "cartel" and "competition", respectively. The algorithm converts the outcome variable through a logistic function to assess the probability of a cartel. We apply the \emph{stats} package by the \citet{team2013r} to implement the logistic regression. 

Our second algorithm, the random forest, is especially suitable to deal with the bias-variance trade-off: With a single classification tree, we can reduce the bias (i.e., systematically wrongly predicting the outcome variable) when we build a "bushy tree" and thus allow for more splits. By doing so, however, we increase the variance (i.e., how much the results vary when the algorithm is applied to different test sets) in the test set due to many leaves. To tackle potential variance in the test set, the random forest builds many trees (in our case, a thousand) by drawing multiple subsamples from the training set. Moreover, to reduce the correlation of trees across these subsamples and thus make the random forest more flexible, the algorithm considers at each splitting step only a subsample of predictor variables, in our case, the square root of the number of predictors. The final predictions consist of the average over all the classification trees. To implement the random forest, we apply the \emph{randomForest} package for the statistical software R by \citet{Liaw2018randomForest}. In addition to the advantages mentioned above, the random forest is relatively user-friendly because it does not require tuning specific penalty terms \citep{athey2019machine}.

Finally, we implement the super learner. This algorithm is a weighted average of multiple machine learning algorithms. We include the random forest \citep{Breiman2001}, lasso logit regression \citep{frank1993statistical,Tibshirani1996}, gradient boosting \citep{friedman2001greedy}, and neural networks \citep{mcculloch1943logical}. The random forest is presented above. The lasso regression is a logistic regression that includes a penalty term, which restricts the number of predictor variables by setting the effect of predictor variables with low predictive power to zero. Similar to the random forest, gradient boosting grows a set of classification trees. In contrast, however, it builds trees in a forward stage-wise manner. The goal is that any additional tree sequentially learns from mistakes made in previous trees. Eventually, a neural network fits a system of nonlinear functions, thereby allowing flexibility in the association between the outcome and the predictor variables. We apply the \emph{SuperLearner} package by \citet{vanderLaanetal2007} together with the \emph{party}, \emph{glmnet}, \emph{xgboost}, and \emph{nnet} packages by \cite{party2022}, \cite{glmnet}, \cite{xgboost}, and \cite{nnet} in the statistical software R.

\subsection{Decentralized approach: A category manager’s tool}\label{Approach2}

In practical use, continuous (automated) data collection in a single structured database may not be feasible. Instead, monitoring of tenders is typically delegated to multiple category managers---so there is a need to equip them with a cartel-screening tool that works decentrally. Category managers' tasks in railway-infrastructure companies are usually quite complex. For example, they must orchestrate construction during operation because the trains usually keep running when the building is in progress. Moreover, they are in constant exchange with planners and construction companies. Therefore, if a railway-infrastructure company expects a category manager to screen the market for potential illegal cartels in addition to daily business, a tool must be as simple as possible (that is, reducing effort cost and enhancing interest in the matter as much as possible). 

\begin{figure}[H] 
	\includegraphics[scale=.4]{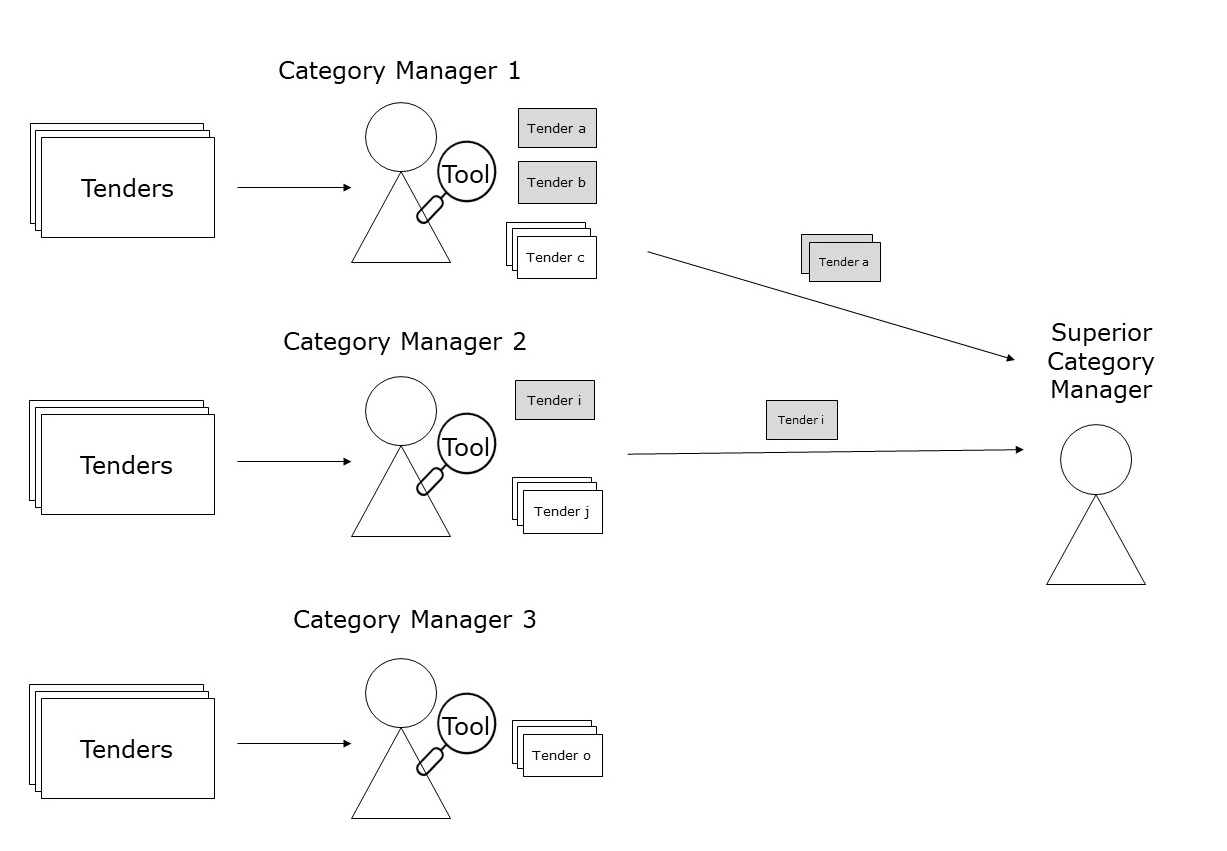}
	\centering \caption{Decentralized approach} \label{Tool1_Fig}
\end{figure}

Figure \ref{Tool1_Fig} depicts the decentralized application of the tool we propose: 
Every category manager is responsible for multiple tenders. He or she investigates each assigned tender immediately using the tool. For instance, Category Manager 1 finds tenders $a$ and $b$ exhibiting suspicious bid patterns (marked grey). He then hands them over to the superior category manager for further investigations. In contrast, the tool does not indicate bid rigging for the tenders of Category Manager 3 and, therefore, no further action is needed.

We propose a threshold-based approach to screen railway-infrastructure markets. Similar to Section \ref{Approach1} and in contrast to theoretically derived benchmarks \citep[see, e.g., ][]{Imhof2019}, we apply a machine-learning algorithm to define the thresholds. More precisely, we suggest the use of a classification (or decision) tree \citep{Breimanetal1984}, a method that is easy to understand (or, at least, easy to utilize) also by non-data-scientists.  In our application, the classification tree uses screens to predict whether a tender is suspicious and, thus, a cartel might be active. The classification of a tender starts at a top node with an associated threshold of the screen. Based on the screen’s value of this tender, a category manager can move down the corresponding branch. The agent repeats this procedure until she arrives at a so-called leaf or terminal node. The terminal node indicates whether the tender is suspicious or not. More technically, we can think of the terminal nodes as non-overlapping regions, i.e., a set of suspicious or non-suspicious tenders. According to a goodness-of-fit criterion, e.g., the Gini coefficient for binary outcomes, the algorithm selects the predictors to maximize the homogeneity of the prediction at each split. 

To prune the tree and to determine the optimal level of tree-complexity, we perform 10-fold cross-validation. For the implementation of the classification tree, we apply the \emph{caret} package for the statistical software R by \citet{kuhn2019}. 

Finally, note that---due to the discrete structure of decision trees---probability thresholds should not be freely chosen in the decentralized approach, which is why we only work with a value of 0.5.

\section{Fitting the models}\label{Fitmodels}

To train our machine-learning models, we draw on data from three convicted bid-rigging cartels provided by the Swiss Competition Commission. These cartels are See-Gaster, Grisons, and Ticino, previously introduced by \citet{Imhof2018} and \citet{wallimann2022machine}.\footnote{See also decisions \textit{Bauleistung See-Gaster: Verfügung vom 8. Juli 2016}, and \textit{Bauleistung Graubünden Strassenbau u.a.: Verfügung vom 19.8.2019} available on the homepage \hyperlink{https://www.weko.admin.ch/weko/de/home/praxis/publizierte-entscheide.html}{https://www.weko.admin.ch/weko/de/home/praxis/publizierte-entscheide.html} (accessed on November 29, 2022). For the Ticino cartel see also on this homepage: \hyperlink{https://www.weko.admin.ch/weko/de/home/medien/medieninformationen/nsb-news.msg-id-16109.html}{https://www.weko.admin.ch/weko/de/home/medien/medieninformationen/nsb-news.msg-id-16109.html} (accessed on November 29, 2022).} The See-Gaster and Grisons cartels were active from 2004 to 2010, while the Ticino cartel was active from 1999 to March 2005. These cartels rigged public and private contracts in construction markets comparable to the railway infrastructure, i.e., road construction, asphalting, and civil engineering. \citet{wallimann2022machine} classify the procurement process in Switzerland as first-price sealed-bid auctions, where all bidders simultaneously submit bids. The bids are "sealed" because bidders do not know their competitors' bids. The price is still the most important---albeit not the single---criterion in determining the winner.\footnote{For instance, even before the overhaul of the Swiss procurement law in 2021, minimum requirements and eligibility criteria were also considered. For more on this, see \hyperlink{https://www.bkb.admin.ch/bkb/de/home/themen/revision-des-beschaffungsrechts.html}{https://www.bkb.admin.ch/bkb/de/home/themen/revision-des-beschaffungsrechts.html} (accessed on February 21, 2023).} Pooling the data of all three cartels leads to 538 competitive (no-cartel) and 519 non-competitive (cartel) tenders. To prevent class imbalances, i.e., preventing the case that one class occurs distinctively more often than the other (a situation where classification methods might not be good at \citep{bekes2021data}), we only consider competitive tenders where past (or future) cartel participants submitted. 

\subsection{Centralized approach}\label{Fit_Approach1}

The screens we use in the centralized approach correspond to the predictor variables described in Appendix \ref{Appendix_A} as well as their interactions and squares.
Table \ref{PerfApproach1} presents the prediction performance in the test set for the three different algorithms (logistic regression, random forest, and super learner). We achieve (decent) correct classification rates from 77.7\% to 81.3\%. In other words, we correctly classify about four out of five tenders. In this regard, the super learner outperforms the two other algorithms by 2.8 percentage points and more. 

\begin{table}[H]
	\begin{center}	
		\caption{Performance measurements in the centralized approach}\label{PerfApproach1}
		\begin{tabular}{lcccc}
			\hline
			\textbf{Algorithm}  & \textbf{CCR} & \textbf{F1 score} & \textbf{95\% PI CCR} & \textbf{95\% PI F1 score} \\ \hline
			Logistic regression & 0.777        & 0.789             & {[}0.717; 0.842{]}                    & {[}0.711; 0.845{]}                         \\
			Random forest       & 0.785        & 0.803             & {[}0.777; 0.857{]}                    & {[}0.773; 0.860{]}                         \\
			Super learner       & 0.813        & 0.796             & {[}0.774; 0.853{]}                    & {[}0.766; 0.889{]}                         \\ \hline
		\end{tabular}
		\par 	{\small \smallskip }
		\textit{Note: PI and CCR denote prediction interval and correct-classification rate, respectively.}
	\end{center}
\end{table}

Since these performance measures might also depend on the test set's composition, we additionally calculate bootstrap prediction intervals by randomly drawing 2,000 training and test sets without replacement. In every training sample, we construct a new prediction algorithm and estimate the performance in the test set. The resulting 95\%-prediction intervals are relatively wide, ranging from 7.9 percentage points (correct-classification rate of super learner) to 13.4 percentage points (F1 score for logistic regression). Therefore, and as a word of caution, the model's performance partially depends on the composition of the training and test sets. However, regarding the random forest and the super-learner algorithms, the lower bound of the interval is still higher than three-quarters correct prediction. 

Concerning 95\%-prediction intervals, we see that the random forest performs almost identically well as the super learner. Therefore, we conclude that both of these algorithms are advisable in the present context. The lower bound of the logistic regression, in turn, is substantially lower (71.7\% for the correct-classification rate and 71.1\% for the F1 score). 

An additional feature of tree-based algorithms is that we can depict the meaningful features. Figure \ref{VI_RF} in Appendix \ref{Appendix_B} shows the relative importance of the predictor variables according to the 2,000 bootstrap samples. We notice that the coefficient of variation is the most important predictor variable. This comes at no surprise, as also other studies consider it as among the most important predictor variables for classifying potential cartels \citep{imhof2021detecting}. 

\begin{figure}[H] 
	\includegraphics[scale=.25]{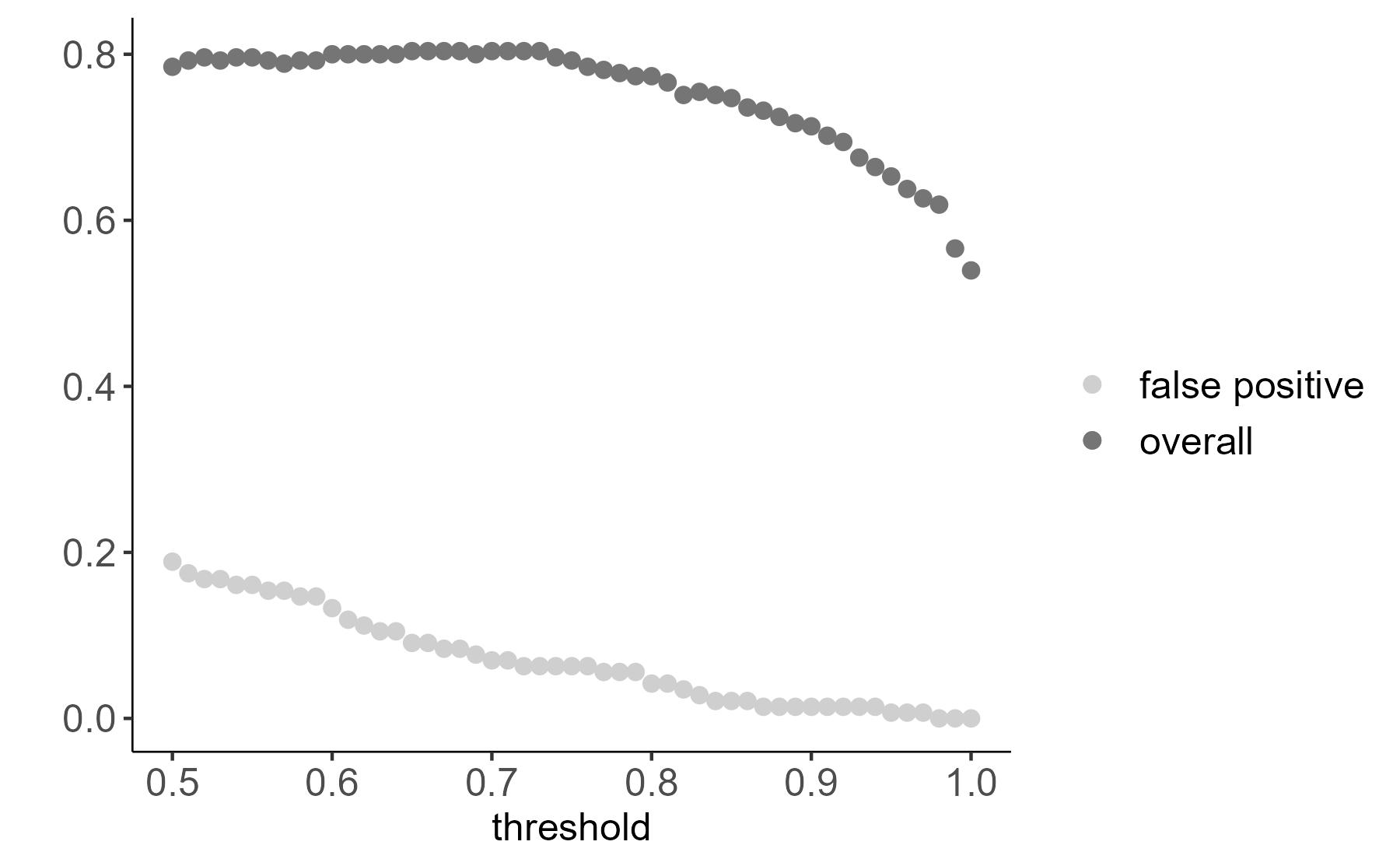}
	\centering \caption{Influence of the probability threshold when applying the random forest} \label{thresholdplot}
\end{figure}

Finally, applying the random forest, Figure \ref{thresholdplot} shows how different decision thresholds affect the performance in the test set: By iteratively increasing the thresholds, the overall correct-classification rate is roughly unaffected up to a threshold of around 0.7, while the false positive predictions substantially decrease (from around 19\% to 8\%). We conclude that even with the higher threshold, we are still able to identify potential cartels. This finding is in line with the thresholds discussed in the literature.

\subsection{Decentralized approach}\label{Fit_Approach2}

The category manager's tool of the decentralized approach consists of a classification tree, which (theoretically, i.e., prior to pruning) utilizes all predictor variables described in Appendix \ref{Appendix_A}. 

Table \ref{PerfApproach2} presents the accuracies of the classification tree in the test set. 

\begin{table}[H]
	\begin{center}	
		\caption{Performance measurements of the category manager’s tool}\label{PerfApproach2}
		\begin{tabular}{lc}
			\hline
			Correct-classification rate                          & 0.800              \\
			F1 score                                             & 0.806              \\
			95\% prediction interval correct-classification rate & {[}0.770; 0.853{]} \\
			95\% prediction interval F1 score                     & {[}0.762; 0.855{]} \\ \hline
		\end{tabular}
	\end{center}
\end{table}

The correct-classification rate amounts to 80\%, the F1 score is about the same. 
Analogous to Section \ref{Fit_Approach1}, we calculate bootstrap prediction intervals, constructing a classification tree for each of 2,000 training samples and estimating the performance in the respective test sets. The resulting 95\%-prediction interval ranges from 77.0\% to 85.3\% for the correct-classification rate. This describes the category manager's tool as similarly predictive as the more opaque algorithms of the centralized approach.

Figure \ref{Classtree} presents the final classification tree for the direct application by category managers. We can easily observe that our classification tree is far from being bushy and has only two terminal leaves. Moreover, only the coefficient of variation (CV) is relevant for classifying a tender as suspicious or not. A category manager can classify a tender as suspicious whenever the CV is lower than 0.053. 


\begin {center}
	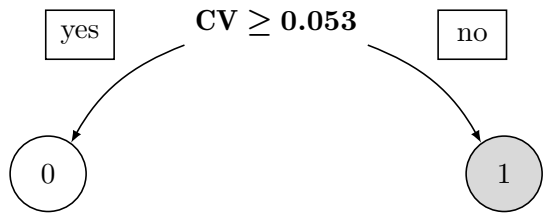
\begin{figure}[H]\centering 
	\begin {tikzpicture}[-latex ,auto ,node distance =2 cm and 3cm ,on grid ,
	semithick]
	\node (CV) {\bm{$\textbf{CV}\geq0.053$}};
	\node [circle, , draw=black, minimum width =1 cm, fill=gray!0, below left=of CV] (0) {0};
	\node [circle, , draw=black, minimum width =1 cm, fill=gray!30, below right=of CV] (1) {1};
	\path (CV) edge [bend left=-20] node[rectangle, draw=black, minimum width=0.9cm, minimum height=0.65 cm, above=0.3cm, xshift=-0.5cm] {yes} (0);
	\path (CV) edge [bend left=20] node[rectangle, draw=black, minimum width=0.9cm, minimum height=0.65 cm, above=0.3cm, xshift=0.5cm] {no} (1);
	\end{tikzpicture}
	\vspace{0.3cm}
	\caption{Classification tree of the category manager's tool} \label{Classtree}
	\end{figure}
\end{center}

\section{Ex-ante application to the Swiss railway-infrastructure market }\label{Application}

In this section, we apply both approaches to real-world data from the railway-infrastructure sector. In particular, we draw on data provided by the Swiss Federal Railways. From a data-analytics perspective, we now use the uncovered patterns from the data of the Swiss Competition Commission and use it to predict the target variable (i.e., a potential cartel's presence or not) in the data of the Swiss Federal Railways. First, we subsequently discuss the institutional background. Then, we present summary statistics of the data set. Finally, we screen the data for collusive behavior using both the centralized and decentralized approaches. We do so by classifying all tenders as “suspicious” and “non-suspicious” in terms of a traffic-light system. The group of suspicious tenders is then analyzed in more detail.

\subsection{Background}\label{Background}

In 2020, the Swiss railway network amounted to 5,317 km.\footnote{See \hyperlink{https://www.bfs.admin.ch/bfs/de/home/statistiken/mobilitaet-verkehr/verkehrsinfrastruktur-fahrzeuge/streckenlaenge.html}{https://www.bfs.admin.ch/bfs/de/home/statistiken/mobilitaet-verkehr/verkehrsinfrastruktur-fahrzeuge/streckenlaenge.html} (accessed on October 12, 2022)} Several railway-infrastructure companies are responsible for building, maintaining, and operating infrastructure. These companies are vertically integrated into state-owned public transport companies. The Swiss Rail Infrastructure Fund (BIF), endowed by general-tax revenue as well as fuel and road taxes, funds expenditures to build, maintain, and operate the railway infrastructure. Since the implementation of the BIF in 2016, an annual amount of 4 to 5 billion Swiss francs has been invested in railway infrastructure, roughly 20\% in building, 66\% in maintaining, and 14\% in operating railway infrastructure \citep{LITRA2022}. The federal transport agency and the railway-infrastructure companies conclude four-year agreements about goals and the distribution of funds to specific tasks to achieve these goals. Based on these agreements and the available funds, the infrastructure companies contract specialized private construction companies for construction projects. Usually, the law requires competitive procedures for contracting out.

\subsection{Data and descriptive statistics}\label{DataDescStats}

We use a unique data set provided by the Swiss Federal Railways, containing projects of track construction, station construction, and civil engineering projects.  The data set contains tenders ranging mainly from 2015 to 2021. Overall, we observe 1,818 tenders. Railway-infrastructure companies in Switzerland can generally choose between two tendering procedures: open procedures and procedures on invitation. Using the latter, a railway-infrastructure company invites a predefined (usually at least three) number of firms. The open procedure does not restrict the participation of firms submitting bids and is thus associated with more severe competition \citep{wallimann2022machine}. 

Note that in order to apply the screens presented above, we must perform some final data wrangling before starting the actual screening. First, since a minimum of three bids is required to calculate the entirety of screens described in Appendix \ref{Appendix_A}, we drop tenders with less than three bids. Second, from time to time, firms submit different variants. In such cases, we only consider the lowest bid. Our final sample contains 1,206 tenders altogether. 

Table \ref{Tenderyear} presents the number of tenders (with more than two bids) per year grouped by the two procedures. For example, there were 267 tenders in 2016, of which 68 belonged to the open procedure and 199 to the procedure of invitation, respectively.

\begin{table}[H]
			\begin{center}	
		\caption{Tenders per year}\label{Tenderyear}
	\begin{tabular}{lccc}
		\hline
		\textbf{}       & \textbf{All} & \textbf{Open procedure} & \textbf{Procedure on invitation} \\ \hline
		older than 2015 & 5            & 0                       & 5                                \\
		2015            & 47           & 21                      & 26                               \\
		2016            & 267          & 68                      & 199                              \\
		2017            & 155          & 46                      & 109                              \\
		2018            & 162          & 31                      & 131                              \\
		2019            & 185          & 49                      & 136                              \\
		2020            & 224          & 56                      & 168                              \\
		2021            & 158          & 38                      & 120                              \\
		2022            & 1            & 1                       & 0                                \\
		Unknown         & 2            & 0                       & 2                                \\ \hline
		Total           & 1,206        & 310                     & 896                              \\ \hline
	\end{tabular}
\end{center}
\end{table}

\begin{figure}[H] 
	\includegraphics[scale=.5]{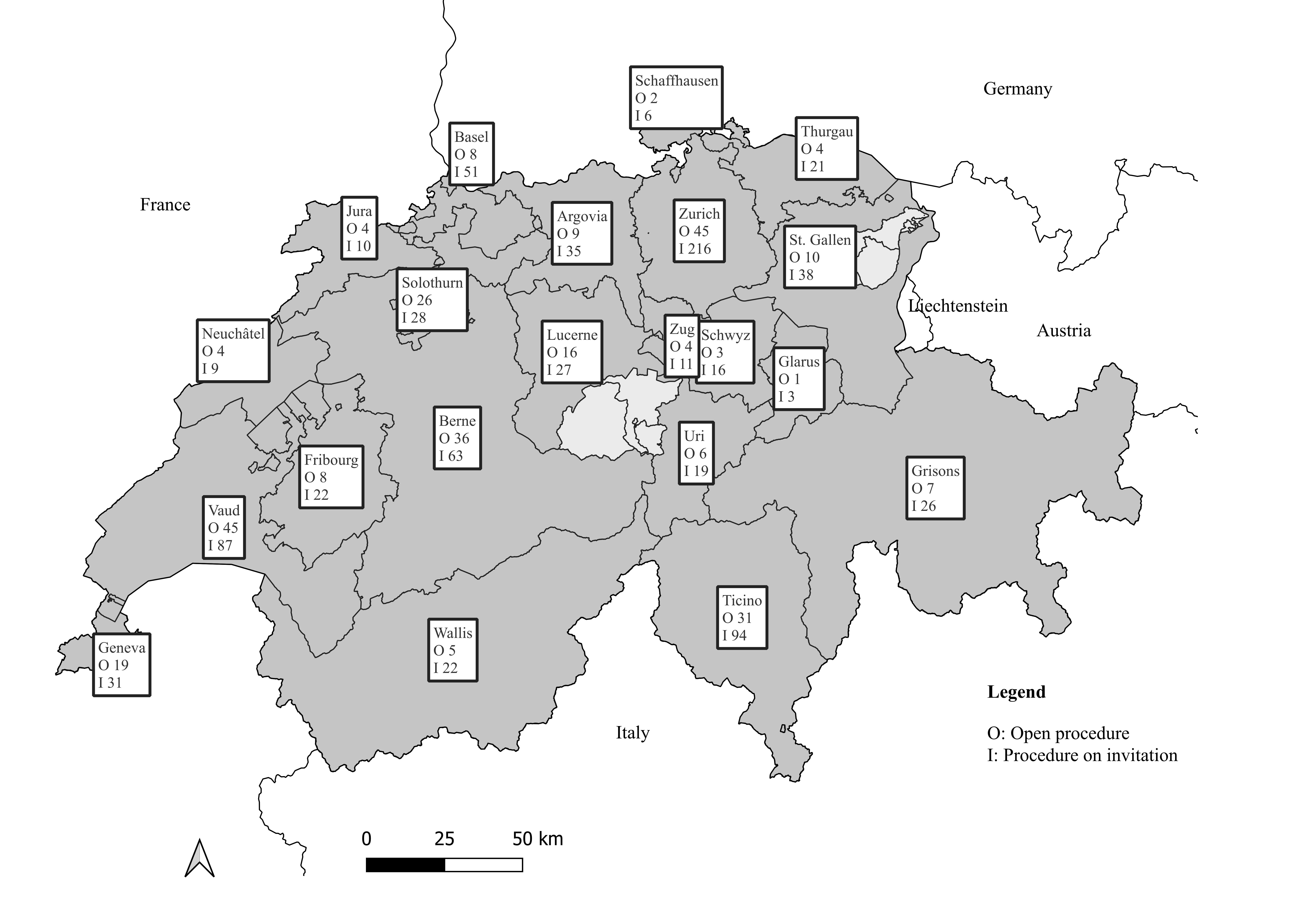}
	\centering \caption{Tenders by canton} \label{Cantons}
\end{figure}

Switzerland consists of 26 member states, the so-called cantons. An interesting feature of our data set is that we obtain data from all over Switzerland. Figure \ref{Cantons} shows the distribution of projects among cantons. We observe, e.g., that the most populous\footnote{With 476\,km, Zurich is also home to the most expansive railway network in Switzerland. See \hyperlink{https://reporting.sbb.ch/infrastrukturen}{https://reporting.sbb.ch/infrastrukturen} (accessed on February 21, 2023).} canton (Zurich) has the most tenders (261), whereby 216 (83\%) and 45 (17\%) are procedures on invitation and open procedures, respectively. The cantons with the second and third highest number of tenders are Vaud and Ticino with 132 and 125, respectively.

Our analysis exclusively rests upon the bids submitted by the firms. Table \ref{Descstats} presents the key descriptive figures of these bids (grouped by tender procedure). We observe, e.g., that the highest bid amounts to approximately 376 million Swiss francs. In comparison, the lowest bid amounts to 2,175 Swiss francs. The mean and median values amount to 2.7 and 0.7 million Swiss francs, respectively, which indicates a heavily right-skewed distribution. We illustrate this distribution also in Figure \ref{bidplot} in Appendix \ref{Appendix_C}. The average bid for projects tendered with the procedure on invitation is lower (0.8 million Swiss francs) than with the open procedures (7.0 million Swiss francs). 

Table \ref{Descstats} also shows that the total number of firms in the data set amounts to 2,061. Of these firms, 1,559 and 779 took part at least once in a procedure on invitation and an open procedure, respectively. 

\begin{table}[H]
	\begin{center}	
			\caption{Descriptive statistics }\label{Descstats}
	\begin{tabular}{lccc}
		\hline
		\textbf{}            & \textbf{All} & \textbf{Open procedure} & \textbf{Procedure on invitation} \\ \hline
		\multicolumn{4}{l}{\textbf{Number of bids}}                                                      \\
		Mean                 & 3.97         & 4.68                    & 3.73                             \\
		Median               & 4            & 4                       & 3                                \\
		\multicolumn{4}{l}{\textbf{Bid values (Swiss francs)*}}                                          \\
		Mean                 & 2,688,906    & 7,021,877               & 804,103                          \\
		Median               & 656,117      & 3,028,289               & 408,199                         \\
		Maximum              & 375,591,800  & 375,591,800             & 36,954,503                       \\
		Minimum              & 2,175        & 11,522                  & 2,175                           \\
		Standard   deviation & 11,180,096   & 19,429,012             & 1,876,232                        \\
		\multicolumn{4}{l}{\textbf{Number of bidding firms}}                                             \\
		Total                & 2,061        & 774                     & 1,559                            \\ \hline
	\end{tabular}
\par
{\small \smallskip }
\begin{spacing}{1}\textit{*Note: We use all bids to calculate the descriptive statistics on the bid values. Therefore, tenders with many bidders have a higher weight.}\end{spacing}
\end{center}
\end{table}

In the following, we illustrate the application of the two approaches discussed in this paper using the data of the Swiss Federal Railways. We present the results in two steps: First, we classify all tenders in terms of a traffic-light rating system as red (i.e., suspicious) or green (i.e., non-suspicious). Second, we screen for potential collusion in subsets of the data (regarding geography, auction format, and firm interactions).

\subsection{Traffic-light rating system and the detection of collusive subsets}\label{TrafficLightSubsets}

Table \ref{Susptend} presents the results of the two approaches' application to the database of the Swiss Federal railways. Applying the random forest and the super learner of the centralized approach with the decision threshold of 0.5, we classify 8.5\% and 5.6\% of all tenders as suspicious, respectively. When we look for "very suspicious" tenders by increasing the decision threshold to 0.7, the random forest and the super learner flag only 2.6\% and 2.4\%, respectively, of all tenders as candidates for further investigations. 

Applying the category manager's tool of the decentralized approach, we see that 130 (10.8\%) of all tenders are classified as suspicious, whereas the vast majority of tenders (89.2\%) again is classified as non-suspicious. 

\begin{table}[H]
		\begin{center}	
		\caption{Absolute and relative amounts of suspicious tenders per algorithm}\label{Susptend}
	\begin{tabular}{lcccc}
		\hline
		\textbf{Approach} & \textbf{Algorithm}  & \textbf{Probability threshold} & \textbf{Suspicious} & \textbf{Non-suspicious} \\ \hline
		\textbf{Centralized}    & Random forest       & 0.5           & 102 (8.5\%)         & 1,104 (91.5\%)          \\
		& Super learner       & 0.5                            & 68 (5.6\%)          & 1,138 (94.4\%)          \\
		& Random forest       & 0.7                            & 31 (2.6\%)          & 1,175 (97.4\%)          \\
		& Super learner       & 0.7                            & 29 (2.4\%)          & 1,177 (97.6\%)          \\ \hline
		\textbf{Decentralized} & Classification tree &           & 130 (10.8\%)        & 1,076 (89.2\%)          \\ 
		\hline
	\end{tabular}
\par
\end{center}
\end{table}

In summary, the results are not indicative of active potential cartels. We conclude that from an overall perspective, procurement of the Swiss Federal Railways is efficient. As expected (regarding the error rates), both approaches classify a subset of several tenders as collusive. This could be due to false-positive results. Nevertheless, if there actually is collusion, it is most likely within theses subsets---rendering them still very worthwhile investigating.

Using the 102 tenders classified as suspicious by the random forest, we investigate whether noticable patterns occor in terms of regions, the procurement process applied, or firm interactions. We use the results of the random forest due to two reasons: First, the centralized approach includes such further analysis. Second, the use of the random forest with the decision threshold of 0.5 has practical reasons, as it identifies more tenders as suspicious than the super learner. Therefore, the chances increase that we find a cluster exhibiting potential cartels.  Moreover, 63 of the 68 tenders marked as suspicious by the super learner also classified as suspicious by the random forest. Therefore, we conclude that the findings of the two algorithms are congruent.

We start by geographically allocating the suspicious tenders to cantons. In Table \ref{Tool2Class05}, we list all cantons with more than 50 tenders. The strongest penetration of suspicious tenders amounts to 14.4\% (present anonymization at the service of the Swiss Federal Railways), which is 5.9 percentage points higher than the nation-wide average. Regarding the two different procedures, we do not see any difference, as for both procedures (open vs. on invitation), the share of suspicious tenders is between 8 and 9 percent (see also Table \ref{Tool2Class05}). Finally, differentiating between years, we see that the amount of suspicious tenders is the highest in 2017. However, with 4.4 percentage points, the difference to the overall amount of suspicious tenders is rather small. 

\begin{table}[H]
	\begin{center}	
			\caption{Screening within clusters (random forest with probability threshold 0.5)}\label{Tool2Class05}
			\begin{tabular}{lcc}
					\hline
					& \textbf{Suspicious} & \textbf{Non-suspicious} \\ \hline
					\textbf{All}            & 102 (8.5\%)         & 1,104 (91.5\%)          \\
					\multicolumn{3}{l}{\textbf{By canton}}                                  \\
					Canton $A$                  & 14.4\%         & 85.6\%             \\
					Canton $B$                    & 13.1\%         & 86.9\%             \\
					Canton $C$               & 9.3\%           & 90.7\%            \\
					Canton $D$                   & 8.5\%           & 91.5\%            \\
					Canton $E$                   & 6.8\%           & 93.2\%            \\
					Canton $F$                  & 4.6\%          & 95.4\%             \\		
					Canton $G$                    & 4.0\%           & 96.0\%            \\		
					\multicolumn{3}{l}{\textbf{By procedure}}                               \\
					Open procedure          & 27 (8.7\%)          & 283 (91.3\%)            \\
					On invitation & 75 (8.4\%)          & 821 (91.6\%)            \\
					\multicolumn{3}{l}{\textbf{By year}}                                    \\
					2015                    & 5 (10.6\%)          & 42 (89.4\%)             \\
					2016                    & 15 (5.6\%)          & 252 (94.4\%)            \\
					2017                    & 20 (12.9\%)         & 135 (87.1\%)            \\
					2018                    & 12 (7.4\%)          & 150 (92.6\%)            \\
					2019                    & 14 (7.6\%)          & 171 (92.4\%)            \\
					2020                    & 21 (9.4\%)          & 203 (90.6\%)            \\
					2021                    & 15 (9.5\%)          & 143 (90.5\%)            \\ \hline
				\end{tabular}
		\end{center}
\end{table}

To investigate the interaction of firms, we present a matrix quantifying how many times firms participated in suspicious tenders together with other firms in Table \ref{Firminteract}. Such a matrix is helpful as we assume that a bid-rigging cartel involves regular interaction between firms \citep{Imhof2018}. Therefore, we only include firms with a minimum of three suspicious tenders and at least one interaction with another firm partaking at three or more suspicious tenders. The diagonal of the matrix contains the percentage of suspicious tenders of each individual firm, where the values in brackets refer to suspicious (bolded) and overall tenders the firm in question took part. For instance, regarding firm 20, 22\% of the tenders (that is, 7 out of 32 tenders) are classified as suspicious. 

\renewcommand{\arraystretch}{2.6}
\begin{table}[t]
	\begin{center}	
		\caption{Firm interaction (random forest with probability threshold 0.5)}\label{Firminteract}
		\footnotesize
		\resizebox{\textwidth}{!}{%
			\begin{tabular}{ccccccccccccc}
				& \textbf{20} & \textbf{227} & \textbf{163} & \textbf{23} & \textbf{179} & \textbf{2411} & \textbf{13} & \textbf{70} & \textbf{89} & \textbf{123} & \textbf{180} & \textbf{273} \\ \cline{2-13} 
				\multicolumn{1}{c|}{\textbf{20}}   & \makecell{22\% \\ (\textbf{7}/32)}   & \makecell{20\%\\ (\textbf{2}/10)}    &              & \makecell{0\%\\ (\textbf{0}/1)}      &              & \makecell{25\% \\(\textbf{1}/4)}      &             &             & \makecell{75\% \\(\textbf{3}/4)}    & \makecell{33\% \\(\textbf{3}/9)}     &              &              \\ 
				\multicolumn{1}{c|}{\textbf{227}}  &             & \makecell{23\%\\ (\textbf{7}/30)}  &              & \makecell{0\% \\(\textbf{0}/1)}     &              & \makecell{25\%\\ (\textbf{1}/4)}      &             & \makecell{67\%\\ (\textbf{2}/3)}    & \makecell{0\%\\ (\textbf{0}/2)}     & \makecell{20\% \\(\textbf{1}/5)}     &              &              \\
				\multicolumn{1}{c|}{\textbf{163}}  &             &              & \makecell{7\% \\(\textbf{5}/72)}     &             & \makecell{4\%\\ (\textbf{1}/26)}     &               & \makecell{0\% \\(\textbf{0}/3)}     &             &             &              & \makecell{0\% \\(\textbf{0}/1)}      &              \\
				\multicolumn{1}{c|}{\textbf{23}}   &             &              &              & \makecell{5\% \\(\textbf{4}/86)}    &              &               & \makecell{7\% \\(\textbf{1}/15)}    &             & \makecell{0\%\\ (\textbf{0}/1)}     &              &              & \makecell{10\% \\(\textbf{1}/10)}    \\
				\multicolumn{1}{c|}{\textbf{179}}  &             &              &              &             & \makecell{7\%\\ (\textbf{4}/59)}     &               & \makecell{0\% \\(\textbf{0}/1)}     &             &             &              &              &              \\
				\multicolumn{1}{c|}{\textbf{2411}} &             &              &              &             &              & \makecell{31\% \\(\textbf{4}/13)}     &             & \makecell{100\% \\(\textbf{1}/1)}   &             & \makecell{0\% \\(\textbf{0}/3)}      &              &              \\
				\multicolumn{1}{c|}{\textbf{13}}   &             &              &              &             &              &               & \makecell{5\% \\(\textbf{3}/63)}    &             &             &              &              & \makecell{13\% \\(\textbf{1}/8)}     \\
				\multicolumn{1}{c|}{\textbf{70}}   &             &              &              &             &              &               &             & \makecell{50\% \\(\textbf{3}/6)}    &             &              &              &              \\
				\multicolumn{1}{c|}{\textbf{89}}   &             &              &              &             &              &               &             &             & \makecell{33\% \\(\textbf{3}/9)}    & \makecell{100\% \\(\textbf{1}/1)}    &              &              \\
				\multicolumn{1}{c|}{\textbf{123}}  &             &              &              &             &              &               &             &             &             & \makecell{18\% \\(\textbf{3}/17)}    &              &              \\
				\multicolumn{1}{c|}{\textbf{180}}  &             &              &              &             &              &               &             &             &             &              & \makecell{17\% \\(\textbf{3}/18)}    &              \\
				\multicolumn{1}{c|}{\textbf{273}}  &             &              &              &             &              &               &             &             &             &              &              & \makecell{9\% \\(\textbf{3}/32)}
		\end{tabular}}
	\end{center}
\end{table}

For the remaining 12 firms, we analyze all $2^{12}=4096$ possible clusters(including trivial clusters with no and one firm). For each of these, we compute a "suspicioucy rate", which we define as the ratio of suspicious to overall tenders involving at least one firm of the cluster. The ranking of firms by suspicioucy rate depends on whether we include the diagonal elements of Table \ref{Firminteract} into our analysis. If we do, the two most "suspicious" clusters are \{70, 89\} and \{70, 2411\}, even though there is only one interaction between firms 70 and 2411 and none between firms 70 and 89. By increasing the cluster size, the suspicioucy rate is barely driven by the interaction of firms. For example, the most suspicious cluster of size four, \{20, 70, 89, 2411\} exhibits a rate of 32\%. However, note that only five out of 22 suspicious tenders relate to interactions.

Once we exclude the diagonal elements, the suspicioucy rate reads as the ratio of suspicious to overall tenders, each involving at least \emph{two} of the firms of a cluster. Still, the top-ranked clusters exhibit little interaction.\footnote{The clusters \{70, 2411\} and \{89, 123\} only interact once, reaching a 100\% rate, because the single interaction is labeled suspicious.} However, further down the list, more interesting cases arise: Cluster \{20, 89\} consists of four interactions, three of them suspicious. Likewise, two out of three interactions are suspicious in cluster \{70, 227\}. Fifth in line, cluster \{20, 70, 89, 2411\} (which we know from above) registers five suspicious interactions out of nine. The thing to notice is that firm 89 never interacts with firms 70 and 2411, and firms 20 and 70 do not interact. Close next is cluster \{20, 89, 123\} with a suspicioucy rate of 50\% (seven out of 14). 

Since cartels (if stable) often and regularly interact, it would be worthwhile to put the companies of these clusters under increased scrutiny in future tenders, even though, up to today, the results are far from striking. 

In summary, we classified between 89.2\% and 97.6\% of all tenders as non-suspicious. These results do not indicate active large-scale cartels. Furthermore, using the results of the centralized approach with the random forest algorithm and a decision threshold of 0.5, we found no striking bid-rigging patterns regarding cantons, procurement procedures, and years. Nevertheless, looking at firm interactions of firms bidding more frequently in suspicious tenders, we identify potential clusters of firms worth increased scrutiny. However, it should be noted that these results may be due to classification errors and should be treated with caution as they are based on small numbers. 

\section{Discussion}\label{Discussion}

Identifying a potential cartel using statistical methods is a thorny task. A suspicious bid pattern in one tender is no definitive evidence for collusion. Nevertheless, it indicates where to look at more closely, or---if no indication of a cartel is found---increases a railway-infrastructure company’s confidence that procurement is competitive. The means we propose for railway-infrastructure companies to deploy exhibit several advantages. First, data requirements are low. This is important as it comes with low effort (for the category manager in the decentralized approach) and allows for cartel screening in large data sets (using the centralized approach). Second, our approaches are relatively simple to implement from a technical point of view. Third, they complement each other. While the centralized approach allows for a systematic overview and various in-depth evaluation options, the category manager's tool of the decentralized approach is particularly well-suited for everyday use. Finally, appropriate communication to the industry helps to deter future collusion in the market. 

Besides these advantages, one should be aware of some limitations. First, the category manager's tool could be seen as too simple since it contains only one benchmark. However, the results presented in section \ref{Fitmodels} demonstrate that it also achieves decent predictions that are almost as accurate as with the centralized approach. Second, being aware of the use of screening tools, firms can try to beat the algorithms. For instance, they could influence the bids in a way that the algorithms classify tenders as "non-suspicious". However, such efforts can be expensive, especially when it comes to the centralized approach, which is a black box for firms. Moreover, in-depth cooperation requires more communication, potentially increasing evidence that a competition agency can use to condemn firms \citep[see also ][]{imhof2021detecting}. Finally, screening tools are just one of several cartel detection methods such as leniency programs. However, increasing the effort to beat algorithms leads to more evidence, making it easier to prosecute potential cartels \citep[see. e.g.,][]{blatter2018optimal}.  

Future studies should investigate how other screening methods could complement the approaches presented in this study.  An example is the application of econometric models to flag potential cartels \citep[see, e.g., ][]{Bajari2003}, as it also takes into account other factors, such as the workload of companies, in order to find potential cartels. One downside of econometric models, however, is their high data requirement, as variables are required that often are intern to firms (such as costs and revenue). Another promising addition could be the application of coalition-based screens \citep{imhof2021detecting}, which screen for groups of suspicious firms instead of suspicious tenders. The main advantage of coalition-based screens is that it directly flags suspicious firms.  However, this is not possible by simple means such as the category manager's tool, which approaches tender by tender. Classifying suspicious groups of firms, in contrast, requires an overview of many tenders in which the same firms submit a bid. Thereby, the coalition-based approach could be promising in identifying collusive subsets when firms are observed frequently interacting in suspicious tenders. 

Moreover, similar to the subsample screening we presented above (regarding geography, different tendering procedures, and firm interactions), practitioners could apply further in-depth analysis---for instance, regarding different types of projects, as some markets could be more susceptible to cartels than others.  

Finally, the external validity of our two approaches is of great practical relevance. External validity means the extent to which the correct classifications are similar in new data sets. For instance, the study of \citet{huber2022transnational} investigates the transnational transferability of screening methods and finds that classification rates can go down when training algorithms in one country and testing performance in another. In our study, external validity refers to whether the patterns of association between the target variable (i.e., potential cartel or competition) and the screens found in the road construction, asphalting, and civil-engineering data apply in railway-infrastructure tenders. We argue that this is the case since the corresponding markets seem comparable due to several reasons: First, the procurement procedures are the same \citep[e.g., ][]{wallimann2022machine}.\footnote{Note, however, that the Swiss Federal Railways chooses the procedure on invitation relatively more often than it is applied in the data set of the Swiss Competition Commission. Therefore, as that influences the number of bidders, we did not consider this variable as a predictor.} Second, in both markets, the companies need heavy machinery and are therefore limited in their radius of action. Third, in both markets, companies depend on comparable suppliers of, e.g., raw materials. Therefore, natural geographical barriers to entry arise from the latter two points. Fourth, we used training data from the same country, i.e., both data sets are from Switzerland. Fifth, we used a broader set of training data (i.e., three cases) to construct prediction models. Finally, an indication is that several engineering and construction firms are also active in the railway-infrastructure market. However, more research on the existence of general cartel patterns across diverse markets is needed. Furthermore, data of upcoming cases has to be used to develop new models able to detect future conspiracies. 

\section{Conclusion}\label{Conclusion}

This study introduced two complementary approaches allow screening for potential cartels by railway-infrastructure companies. Both approaches rely on machine-learning algorithms, which assess whether tenders show suspicious or non-suspicious patterns in terms of potential collusion among bidders. We draw on previous investigations of the Swiss Competition Commission to uncover such patterns. We used these patterns to predict potential cartel behavior in railway-infrastructure procurement by means of a traffic-light system. 
	
First, we applied several state-of-the-art machine-learning algorithms to flag collusive agreements of firms. The centralized approach of doing so allows a systematic overview of a railway-infrastructure company's tenders and various in-depth analyses.  

Second, we presented a tool tailored to individual category managers responsible for projects put out to tender. When a category manager detects bids exhibiting suspicious patterns, she contacts the superior category manager for further investigations. 	

Be it by actually disclosing cartels or by deterring their formation, the endeavor of screening tenders is of high practical relevance in light of the vast amounts of (public) money devoted to railway infrastructure. Finally, the thing to notice is that the two approaches can also serve as examples for uncovering potential cartels in other industries.

	\newpage
	\bigskip
	
	\bibliographystyle{econometrica}
	\bibliography{SellerTrackStations.bib}
	
	\bigskip
	
\begin{appendix}
		
		\numberwithin{equation}{section}
		\counterwithin{figure}{section}
		\noindent \textbf{\LARGE Appendices}
	
	\section{Formally introduction of the screens}\label{Appendix_A}

	To formally introduce the screens note that $t$ denotes some tender $t$; ${\bar{b}_{t}}$ is the mean of all bids in tender $t$; $sd_{t}$ and $d_{losing bids,t}$ are the standard deviation of all bids and the loosing bids not including the lowest one of tender $t$ respectively; $n_{t}$ is the number of bids submitted in tender t; $b_{max,t}$ is the highest bid of tender $t$ respectively; $b_{i,t}$ and $b_{j,t}$ are the $i-th$ and the $j-th$ bids of tender $t$, which are ordered from lowest to highest bids; $i_{t}$ is the rank of bid $i$. We calculate the screens using the following formulas: 

	\begin{equation}\label{eqcvMLS}
		CV_{t}=\frac{sd_{t}}{\bar{b}_{t}},
	\end{equation}

	\begin{equation} \label{SPD}
	SPD_{t}=\frac{b_{max,t}-b_{1,t}}{b_{1,t}},
	\end{equation}

	\begin{equation} \label{DiffPerMLS}
		DIFFP_{t}=\frac{b_{2t}-b_{1,t}}{b_{1,t}},
	\end{equation}

	\begin{equation} \label{RDTMLS}
		RD_{t}=\frac{b_{2,t}-b_{1,t}}{sd_{losing bids,t}},
	\end{equation}

	\begin{equation} \label{ALTRDTMLS}
		RDALT_{t}=\frac{b_{2,t}-b_{1,t}}{\frac{(\sum_{i=2,j=i+1}^{n_{t}-1}b_{j,t}-b_{i,t})}{n_{t}-2}},
	\end{equation}

	\begin{equation} \label{RDNORMTMLS}
		RDNOR_{t}=\frac{b_{2,t}-b_{1,t}}{\frac{(\sum_{i=1,j=i+1}^{n_{t}-1}b_{j,t}-b_{i,t})}{n_{t}-1}},
	\end{equation}

	\begin{equation} \label{kolmostat}
		KSTEST_{t}=max(D_{t}^{+},D_{t}^{-}) \mbox{ with } D_{t}^{+}=max_{i}(\frac{b_{i,t}}{sd_{t}}-\frac{i_{t}}{n_{t}+1}), D_{t}^{-}=max_{i}(\frac{i_{t}}{n_{t}+1}-\frac{b_{i,t}}{sd_{t}})
	\end{equation}

	\section{Training the algorithms}\label{Appendix_B}

		\begin{figure}[H] 
		\includegraphics[scale=.25]{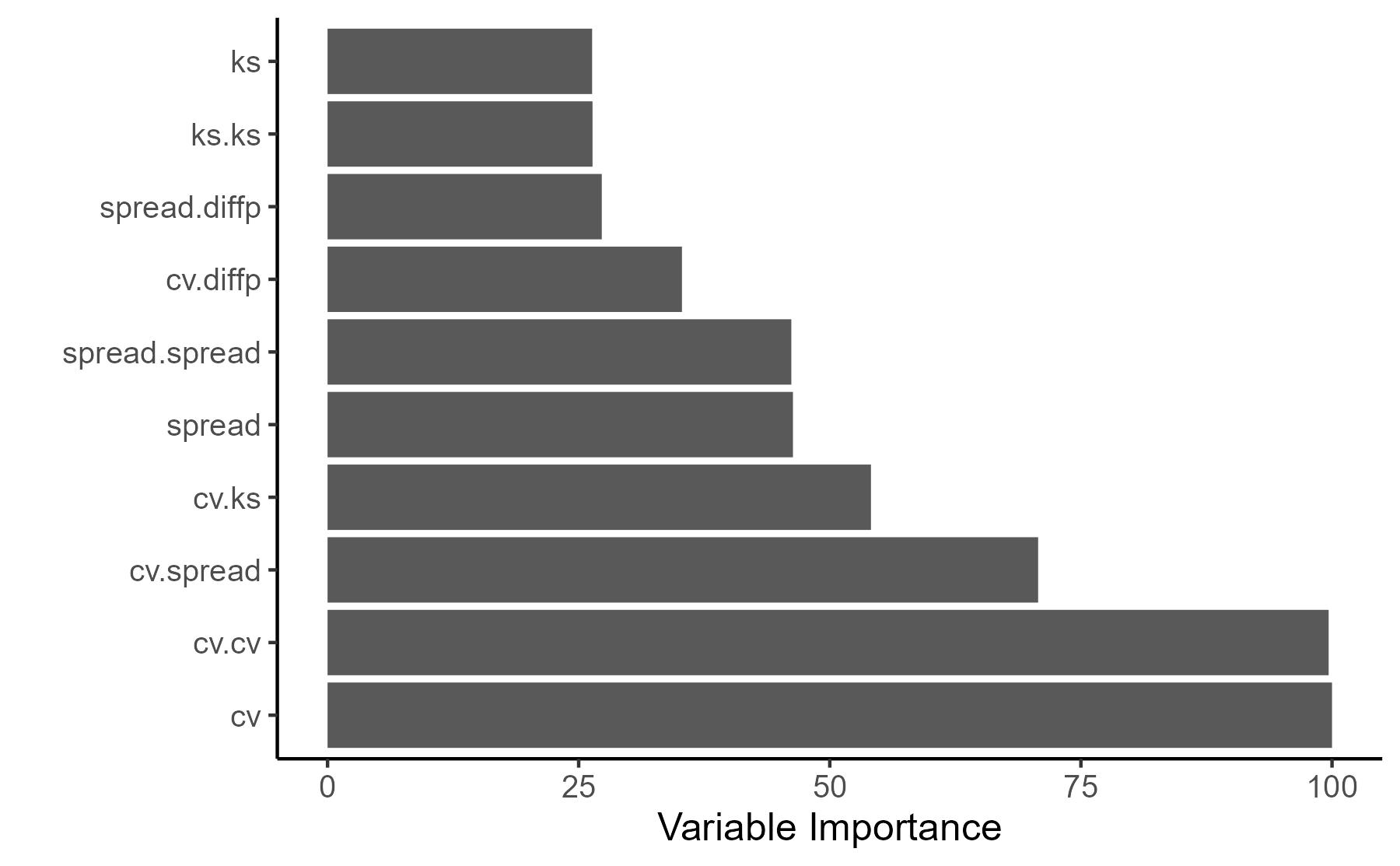}
		\centering \caption{Variable importance plot of the random forest. Note, we express the importance of predictor variables relative to the maximum.} \label{VI_RF}
	\end{figure}

	\section{Additional Figures and Tables}\label{Appendix_C}
	
	\begin{figure}[H] 
		\includegraphics[scale=.25]{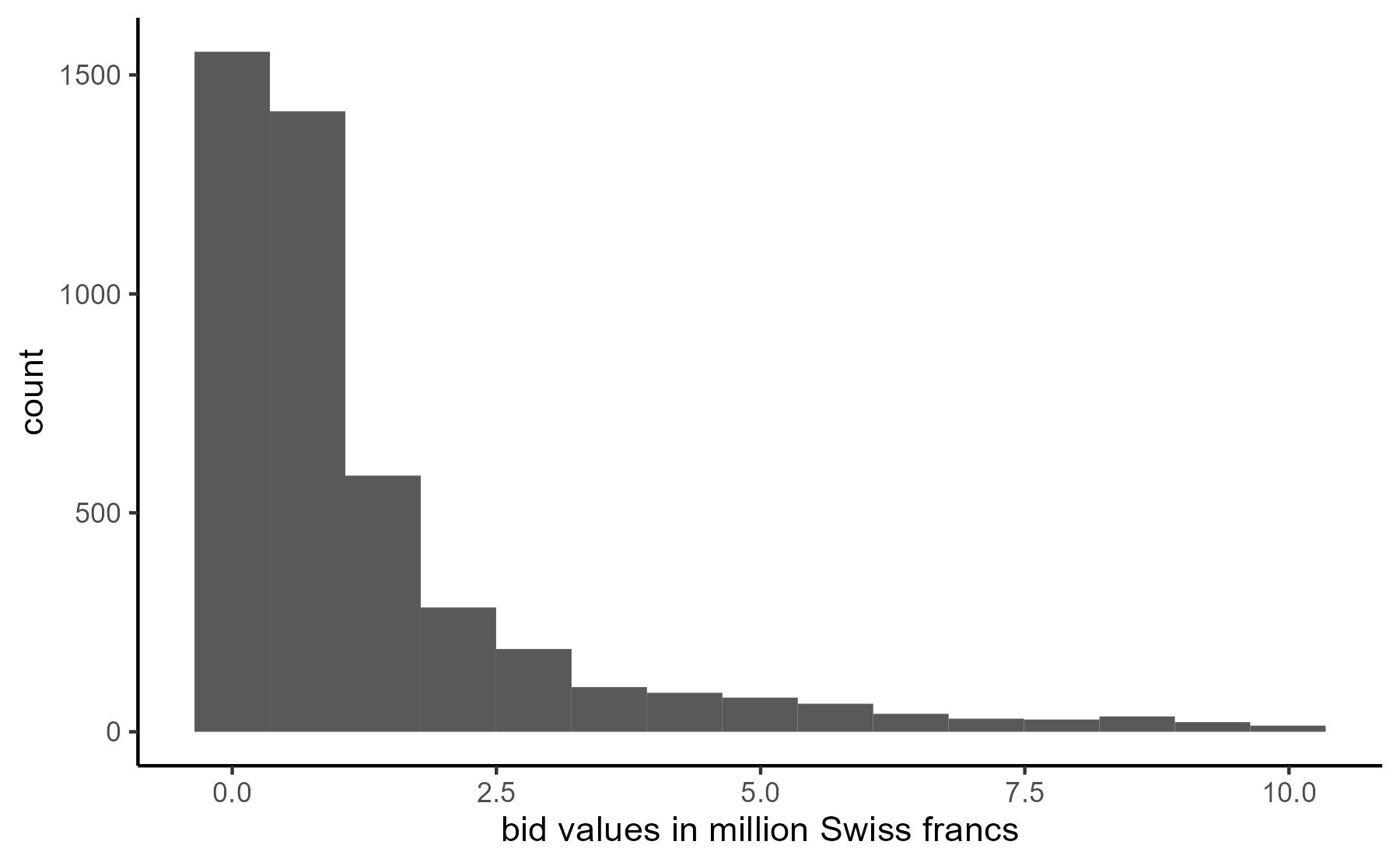}
		\centering \caption{Bid values in million Swiss francs (note: bid values higher 10 million Swiss francs, i.e., 5.4\% of all bids, have been removed for the plot)} \label{bidplot}
	\end{figure}

	\end{appendix}
\end{document}